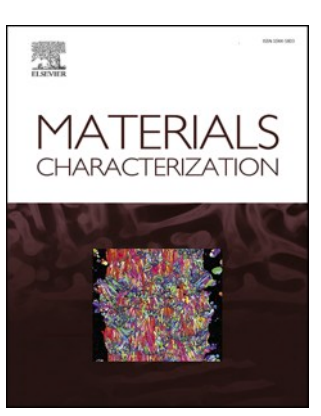

# Human perception-inspired grain segmentation refinement using conditional random fields

Doruk Aksoy [a,*], Huolin L. Xin [b], Timothy J. Rupert [a,c], William J. Bowman [a,*]

[a] Department of Materials Science and Engineering, University of California, Irvine, CA 92697, USA
[b] Department of Physics and Astronomy, University of California, Irvine, CA 92697, USA
[c] Department of Mechanical and Aerospace Engineering, University of California, Irvine, CA 92697, USA

ABSTRACT

Automated detection of grain boundaries in electron microscope images of polycrystalline materials could help accelerate the nanoscale characterization of myriad engineering materials and novel materials under scientific research. Accurate segmentation of interconnected line networks, such as grain boundaries in polycrystalline material microstructures, poses a significant challenge due to the fragmented masks produced by conventional computer vision algorithms, including convolutional neural networks. These algorithms struggle with thin masks, often necessitating post-processing for effective contour closure and continuity. Previous approaches in this domain have typically relied on custom post-processing techniques that are problem-specific and heavily dependent on the quality of the mask obtained from a computer vision algorithm. Addressing this issue, this paper introduces a fast, high-fidelity post-processing technique that is universally applicable to segmentation masks of interconnected line networks. Leveraging domain knowledge about grain boundary connectivity, this method employs conditional random fields and perceptual grouping rules to refine segmentation masks of any image with a discernible grain structure. This approach significantly enhances segmentation mask accuracy by correctly reconstructing fragmented grain boundaries in electron microscopy images of a polycrystalline oxide. The refinement improves the statistical representation of the microstructure, reflected by a 51 % improvement in a grain alignment metric that provides a more physically meaningful assessment of complex microstructures than conventional metrics. This method enables rapid and accurate characterization, facilitating an unprecedented level of data analysis and improving the understanding of grain boundary networks, making it suitable for a range of disciplines where precise segmentation of interconnected line networks is essential.

## 1. Introduction

Integrating computer vision with electron microscopy has significantly enhanced efficiency in materials science by speeding up traditionally laborious and time-consuming processes [71]. The advancement of (scanning) transmission electron microscopy ((S)TEM) is pivotal in examining grain boundary (GB) networks and other microstructural features in hard materials like metals and ceramics [72]. These techniques yield insights into structural characteristics such as (dis)order [75], dislocations [56], phase transformations and GB segregation [53]. GB networks, often statistically represented [74], require extensive data for analysis. Revealing large-scale information like average grain sizes facilitates high-throughput experiments [80] aimed at elucidating local GB properties [72] and atomic (dis)order [75]. However, developing these techniques is challenging due to the complex nature of grain structures and their boundaries in ceramics, metals, and composites [68,69], involving issues like managing overlapping grains [49] and deciphering defects of different dimensionality such as GBs, triple junctions, and nodes [6]. This complexity is accentuated when analyzing GB networks, influenced by factors like crystallographic orientations, interfacial segregation, and varied local atomic environments, affecting the mechanical, electrical, chemical, and magnetic behavior of materials [1–3,5,10–13,37,51,63,78]. GBs range from ordered high-symmetry structures to high-energy disordered configurations [24], with networks exhibiting intricate relationships due to energetic competition between interfacial planes, atomic sites, and solute-solute interactions [4,26,70]. Moreover, local ordering and local hardening can be influenced by solute segregation to GBs [6, 45].

* Corresponding authors.
*E-mail addresses:* doruka@gmail.com (D. Aksoy), will.bowman@uci.edu (W.J. Bowman).

Therefore, developing predictive techniques is challenging, considering the multifaceted role of GBs where local atomic environments, segregation behaviors, and solute interactions are closely connected. This complexity often results in (S)TEM image contrasts, posing a unique challenge for conventional segmentation methods in identifying GBs.

Addressing this, semantic segmentation – a computer vision strategy categorizing each pixel into a class [46] – emerges as a promising approach for autonomous grain segmentation. Historically, two techniques have been prominent: traditional image processing and modern computer vision techniques. Traditional methods focus on low-level image details, using techniques like thresholding [54], morphological processing [21], edge detection with preset filters [29], unsupervised machine learning clustering [73], watershed transformation [62], and region-growing [61]. Their effectiveness often depends on image quality, influenced by factors like resolution, color balance, brightness, and gradient similarity [44]. Conversely, modern computer vision approaches, particularly convolutional neural networks (CNNs), provide a more sophisticated, high-level grain representation [23,27,55,58]. Architectures like DeepLab [18], Mask R-CNN [30], and U-Net [60] are employed for their adaptability to different image resolutions and consistent segmentation capabilities.

However, CNNs' large perception fields limit their ability to produce accurate pixel-level labels [46], which poses a problem in an interconnected grain network where the segmentation mask is a few pixels wide with a label assigned to each pixel (e.g., 1: GB, 0: grain interior). Metrics like intersection-over-union (IoU) and Dice similarity coefficient (DSC) are used to assess the alignment of predicted masks with ground truth. But, in thin masks, even minor 1-pixel deviations significantly affect accuracy [19]. Researchers have explored various categories of post-processing methods to enhance the segmentation accuracy of thin, interconnected line structures like GBs. One prominent category involves probabilistic models, which refine initial segmentation masks by enforcing spatial consistency. For instance, one such approach has been successfully applied to road network extraction from aerial imagery by smoothing fragmented predictions and maintaining continuity [9]. This was achieved through an iterative search process, where a CNN-driven decision function directly generated the road network graph from the CNN's output. In another application, Dulau et al. [25] demonstrated this by improving the continuity of retinal blood vessel maps, by removing misclassified pixels erroneously identified as retinal vessels, and by rejoining vessel segments that are correctly identified yet remain separated. However, these techniques typically rely on heuristic rules and can struggle with noisy or incomplete initial predictions.

Morphological and perceptual grouping approaches represent another prevalent set of post-processing methods. Morphological operations, such as dilation and skeletonization, have been widely used for tasks like cleaning up segmentation masks or connecting small gaps in biomedical vessel segmentation [40]. Perceptual grouping methods, like tensor voting, have been employed effectively to reconnect fragmented segments in pavement crack detection [81]. Despite their utility, these methods often require extensive manual tuning of parameters, limiting their general applicability across diverse datasets. In contrast, model-specific integrated post-processing methods, like Li et al. [42], have demonstrated remarkable accuracy by incorporating generative adversarial networks and multi-task learning explicitly designed for GB segmentation. Furthermore, Patrick et al. [52] introduced an automated grain boundary detection framework based on the U-Net architecture for BF TEM images that achieves high segmentation accuracy with minimal manual intervention, improving processing throughput. Yet, the specialized nature and reliance on tailored parameters significantly restrict their general use beyond specific materials or image types. Main challenges include connecting adjacent pixels and removing isolated ones, relying on user-defined thresholds. These methods often miss crucial constraints related to GBs, essential for accurate segmentation of complex grain structures.

Here, this work aims to address the gap in semantic segmentation for (S)TEM images of polycrystalline materials by introducing a versatile post-processing method to refine interconnected GB network masks. This method, based on insights into GB connectivity [43] and the interconnected nature of GB networks [14], is designed for broad applicability across different crystalline materials. At its core, it employs segmentation masks from computer vision models to generate conditional probability maps using conditional random fields (CRFs), bridging gaps left by CNNs. The required feature functions for CRFs are selected based on perceptual grouping rules [50]. Applicable to any crystalline material with discernible grain structure, and adaptable to other domains with interconnected line networks, this method can be used as a post-processing step for various grain segmentation masks from imaging systems and vision algorithms, enabling real-time segmentation refinement. It marks a significant advancement for materials scientists, allowing rapid and precise GB segmentation in complex microstructures with poorly resolved boundaries, and enabling previously inaccessible large-scale data analysis, greatly enhancing statistical representation of GB networks.

## 2. Methods

The systematic approach for grain segmentation refinement, aimed at addressing the limitations of conventional computer vision algorithms, is outlined in the subsequent sections. Initially, a computer vision model is developed, serving as a preliminary step to generate segmentation masks. These thin segmentation masks which include fragmented sections form the foundation for applying the proposed hierarchical CRF method. This method, along with perceptual grouping principles, is introduced to address various aspects of grain boundary properties and connectivity. For a given image, the solution that maximizes the conditional probability obtained from the hierarchical CRF corresponds to a label configuration that determines points of connection. Subsequently, a novel path tracing algorithm is introduced, designed to connect fragmented segments using the label configuration derived from the CRF predictions.

### 2.1. Computer vision model development

The input dataset for model training consisted of high-angle annular dark field and annular bright field STEM images recorded with a Nion UltraSTEM100 aberration-corrected STEM operating at 60 kV. The material is an electrically conducting polycrystalline oxide, fluorite $Pr_{0.1}Ce_{0.9}O_2$, synthesized by pulsed laser deposition as a layer of nearly uniform thickness (~30 nm) atop an amorphous silicon nitride free standing substrate.

The images in the dataset are digitized and manually annotated by tracing GBs by hand using a digital drawing tablet, a process widely utilized in materials science, to assess grain sizes and network structures for subsequent analysis of material properties. GBs marked on the images constitute the computer vision ground truth (CVGT), serving as targets for model training. To minimize human operator bias, both annular dark field and bright field images of the same areas were annotated at different times. Regions marked as interfaces in both image types reinforce the trained model's predictions of interface pixels.

A pre-processing pipeline is implemented to enhance data diversity and simulate realistic experimental conditions. The dataset is initially partitioned at the level of the original images into training and validation sets using a 70–30 split. Each image is subsequently subdivided into non-overlapping patches of 512 × 512 pixels. Each patch typically contains approximately 64 grains, capturing the inherent repetitive microstructural features of the material.

To further improve model generalization and simulate experimental variability, an augmentation pipeline is applied exclusively to the training patches. This pipeline performs random brightness adjustments (factors ranging from 0.8 to 1.2), Gaussian blurring (with sigma values between 0.5 and 1.5), and rotations (between −45° and 45°). By applying 20 random augmentations per training patch, the effective number of training samples is expanded to 4480, while the validation set remained unaltered at 144 patches to ensure an unbiased evaluation of model performance.

A computer vision algorithm for image segmentation is developed using a modified U-Net architecture [60], with VGG16 model [66] serving as the encoder backbone. This architecture, effective for varying image resolutions, features a symmetric encoder-decoder structure [60] with skip connections to preserve key details lost during downscaling [31]. It consists of a contracting path capturing low-level features while reducing spatial size and increasing channels, and an expanding path that upscales these features to original size, reducing channels and capturing high-level features for accurate segmentation [60]. The model includes four decoder blocks, each with a 2D convolution layer matching the corresponding encoder layer's channels, followed by concatenation with the encoder's skip connection and two convolutional blocks with batch normalization and ReLU activation. The final output uses a sigmoid function. Compiled with Adam optimizer [36] with a learning rate of $10^{-4}$ and a custom loss function $(1 - DSC)$, the model employs early stopping based on validation loss improvement, monitoring metrics like DSC, IoU and binary accuracy.

The objective of this work is not to achieve the highest possible segmentation accuracy but rather to demonstrate that the post-processing model is learning meaningful features rather than overfitting. To this end, no extensive hyperparameter tuning or further optimization techniques, such as learning rate scheduling, ensemble methods, or advanced regularization, were pursued. Instead, a 3-fold cross-validation strategy is employed on the augmented training set to robustly assess the model's performance. This approach ensures that the model achieves a good enough performance level, with stable validation results indicating that it is effectively capturing the key features of the GBs without overfitting.

### 2.2. A hierarchical CRF architecture

When labeling sequence is critical, such as for interface pixels, CRFs emerge as a versatile tool. As a discriminative model, CRFs calculate the conditional probability of a pixel or segment having the same label given a sequence of observations [67]. CRFs require feature functions, mathematical representations of feature relationships [67]. An essential characteristic of CRFs, rooted in their Gibbs distribution formulation, is the concept of conditional independence. This implies that two pixels or segments without a direct interaction (as defined by the CRF graph structure) become conditionally independent once the state of their neighborhoods is fixed. Thus, distant pixels, though possibly globally correlated through intermediate connections, are directly independent in the conditional sense. Practically, this means the CRF explicitly models local and selected long-range interactions, making computational complexity manageable while preserving meaningful global coherence.

To incorporate domain knowledge accompanying the CRF method, the feature functions can be selected based on perceptual grouping principles. These principles, derived from Gestalt laws of perceptual organization, describe how humans intuitively perceive and interpret the visual world [50] and are relevant to grain segmentation refinement, as illustrated in Fig. 1. Key principles include:

- The Law of Proximity, indicating that elements close together are perceived as a group. In Fig. 1a, clustered dots exemplify this, aiding in identifying nearby pixels or segments as part of the same GB, especially areas of closely packed or intersecting boundaries.
- The Law of Similarity, where similar elements are seen as related. Fig. 1b shows this with dots of the same color and shape, grouping pixels or segments with similar characteristics for continuous GB identification.
- The Law of Closure, involving the mind's tendency to complete incomplete figures, as in Fig. 1c. This helps in filling gaps in GB

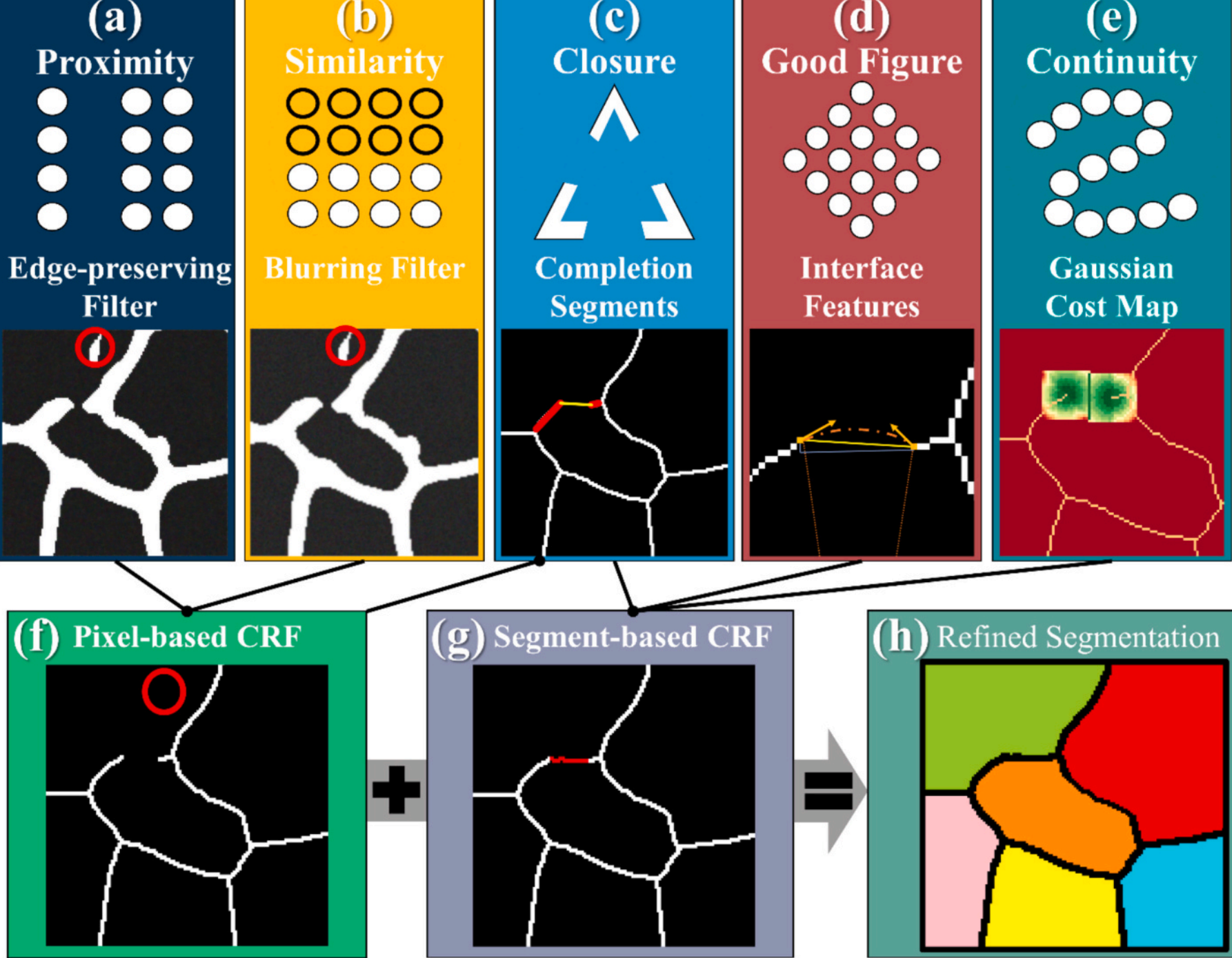

**Fig. 1. Overview of the utilization of the Gestalt principles of perceptual grouping**. (**a**) Proximity, (**b**) similarity, (**c**) closure, (**d**) good figure, and (**e**) continuity principles are satisfied through the use of (**f**) pixel-based CRF, and (**g**) segment-based CRFs resulting in (**h**) a refined segmentation. Each subplot provides a distinct visualization or method that contributes to the overarching theme of the paper, the robust analysis of GB networks.

networks that appear broken, addressing challenges posed by poorly resolved and complex structures in TEM images.

- The Law of Good Figure (or Prägnanz), where complex images are perceived in their simplest form, as seen in Fig. 1d. This aligns with GBs' tendency to minimize energy by adopting simpler shapes, reflecting the natural formation and evolution of GBs in materials.
- The Law of Continuity, favoring smooth patterns. Fig. 1e demonstrates elements aligned in a line or curve as a cohesive group, essential for maintaining continuous and smooth paths consistent with the physical properties of grain structures.

Fully-connected independent pixelwise classification, often leading to patchy segmentation masks due to disregarding pixel relationships, is computationally demanding [8]. CRFs are preferred in such contexts as they effectively incorporate spatial relationships between pixels using pairwise costs. Early models primarily utilized grid graphs for defining these costs, which restricted information transfer among adjacent pixels [65]. However, the introduction of DenseCRF architectures allowed for fully connected graphs and more expressive pairwise potentials [22,38]. Higher-order CRFs for contour completion bring three key benefits to grain segmentation: (i) they foster label continuity among spatially and intensity-wise close pixels, (ii) facilitate modeling of long-range interactions, and (iii) use linear inequalities for modeling extension and continuity constraints.

In this work, a hierarchical methodology for object recognition and segmentation is introduced, combining pixel-based and segment-based definitions synergistically. The novelty of this approach lies in its hypergraph architecture, which integrates a DenseCRF for low-level pixel features with a higher-order, segment-based CRF that enforces domain-specific knowledge through perceptual grouping rules. This synergistic combination allows the model to refine fragmented boundaries by considering both local pixel evidence and global structural coherence, a capability not explicitly present in standard DenseCRF or higher order CRF models alone. The architecture is designed as a hypergraph, with vertices as individual pixels and hyperedges as segments. Inter-segment connections are construed as edges between hyperedges. Initially, a DenseCRF architecture [38] processes low-level pixel features (Figs. 1(a), (b), and (f)), which then inform the higher-level segment-based features, as illustrated in Figs. 1(c)-(e), and (g). This layered strategy allows for the integration of both local and global information, adhering to perceptual grouping principles.

The conditional probability of segments, which describes the probability distribution over all possible label configurations, is determined by minimizing Gibbs energy, comprising unary, pairwise, and higher-order terms. The unary potential encodes the likelihood of individual pixels belonging to specific segments, either complete or broken. This potential is derived from the pre-trained CNNs that provide scores for each pixel, essentially constituting an initial coarse-grained classification. Local consistency is subsequently enforced through edge-preserving bilateral and Gaussian blur filters, yielding perceptually coherent regions by evaluating neighboring pixels' features like color and proximity, as shown in Figs. 1(a) and (b). This effectively distinguishes segments in densely packed GBs, with non-conforming pixels eliminated, depicted in Figs. 1(a), (b), and (f) with red circles. Subsequently, a graph-based technique is employed to identify completion segments, which represent potential connections between broken segments. Due to computational limitations, only a subset of these completion segments, selected based on predetermined criteria, are deemed viable segments for further analysis. The various segment types are visually differentiated in Fig. 1c, with complete, broken, and completion segments shown in white, red, and yellow, respectively.

The pairwise potential, designed to capture long-range interactions, functions effectively on identified segments to maintain global consistency. It enforces completion and extension constraints, as described in a previous contour completion model [79], aligning with the principle of closure illustrated in Fig. 1c. The completion constraint activates a completion segment only when its adjacent broken segment is active. Simultaneously, the extension constraint ensures that an active broken segment has at least one neighboring active completion segment, preserving the continuity and integrity of the GB structure.

A higher-order potential incorporates specific visual characteristics like contour sharpness, smoother segment transitions, angular differences between segments, and segment lengths. The model's complexity is managed by considering additional features such as the total effective length of connected segments and the angular deviation from ideal connection angles. This optimizes simpler, more stable GB forms, adhering to the law of good figure, reflecting the natural evolution of GBs to minimize energy and simplify complex structures for perceptual coherence, as shown in Fig. 1d.

In this composite architecture, the hypergraph method efficiently captures intricate relationships between pixels and segments, enabling a thorough representation of spatial and geometric features in the image. The primary goal is to achieve a binary label configuration that maximizes the conditional probability, representing the CRF's maximum a posteriori probability estimate. This poses a combinatorial optimization challenge, where the ideal solution is the label configuration maximizing this probability. This optimization problem can be posed as a mixed-integer linear programming (MILP) problem, which can then be solved to obtain the optimal label configuration.

Identifying the optimal label configuration involves determining local weights for interface features and global weights for unary, pairwise, and higher-order potentials. The optimal configuration is that which most closely approximates the ground truth, leading to an NP-hard combinatorial optimization problem. Being NP-hard, it is defined by the property that while the validity of a solution, like a label configuration, can be efficiently verified via the MILP equation, there is no polynomial-time algorithm for optimal resolution. Consequently, exact solutions to such NP-hard problems are computationally infeasible [77], especially in images with hundreds of broken and completion segments, typical of fragmented GB networks.

To address this challenge, this study employs differential evolution to minimize the objective function, achieving a near-optimal solution. As an evolutionary algorithm and global optimization method, differential evolution iteratively refines candidate solutions, adjusting their weights based on performance metrics within set bounds [57]. This process derives weights for the optimal label configuration by repeatedly solving the MILP with varied weight combinations. The resulting optimal label configuration, indicated by activated segments (marked as 1 by the CRF), facilitates the smooth connection of completion segments. This approach aligns with the law of continuity, essential for ensuring continuous, logically flowing GBs in the segmentation process, consistent with the physical nature of grain structures, as depicted in Fig. 1e.

### 2.3. Applying the hierarchical CRF to probabilistic image labeling

The conditional probability, $P(Y|X)$, describes the probability distribution over all possible label configurations $Y$ given the observed data $X$. $X$ refers to the entire set of observed data over the complete image, containing all of the features (e.g., pixel intensities, color channels, spatial segment information and any other characteristics of the images) that are used to determine the labels $Y$. $Y$ refers to the set of all labels or the complete labeling configuration for an entire image, that is represented as a vector containing the labels for each pixel or segment. On the other hand, $y$ refers to the label of an individual unit, such as a single pixel or segment. Essentially, each $y_i$ is an element of the set $Y = \{y_1, \ldots, y_n\}$. $Y_{p_i}$ and $Y_{s_i}$ refer to the label configurations assigned to pixels $p_i$ and segments $s_i$ that are part of a set, respectively. The indices $i$, $j$, and $k$ represent the different entities in the image, where $i$ is the pixel or segment that is currently being labeled, and $j$ and $k$ represent other pixels or segments in the same clique as $i$. The relationships between $i$, $j$, and $k$ are encapsulated in the pairwise and higher-order terms of the CRF, capturing local and higher-order interactions in the image. The overall probability distribution $P(Y|X)$ is defined as:

$$P(Y|X) = \frac{1}{Z(X)}\exp\{-E(Y|X)\}, \tag{1}$$

and

$$E(Y|X) = E_u\left(Y_{s_i}\right) + E_p\left(Y_{s_i}, Y_{s_j}\right) + E_h\left(Y_{s_i}, Y_{s_j}, Y_{s_k}\right), \tag{2}$$

where $E_u(Y_{s_i})$, $E_p(Y_{s_i}, Y_{s_j})$, and $E_h(Y_{s_i}, Y_{s_j}, Y_{s_k})$ are the unary, pairwise, and higher-order Gibbs energies, respectively. These energies are calculated based on the labeling of segments $s_i$, $s_j$, and $s_k$, not just individual pixels. $Z(X)$ is a partition function that ensures a normalized probability distribution (i.e., between 0 and 1), given by:

$$Z(X) = \sum\nolimits_{s_i,s_j,s_k} \exp\{-E(Y|X)\}. \tag{3}$$

The unary energy term in Eq. 2, $E_u(Y_{s_i})$, is derived from the pixel-based CRF where the unary term is generated by a CNN and the pairwise term is obtained from a Gaussian kernel. Explicitly, for the unary term, the segment-based $E_u(Y_{s_i})$ can be defined as:

$$E_u\left(Y_{s_i}\right) = \Omega^u \mathscr{C}\left(\sum\nolimits_{p_i\in Y_{s_i}} \psi_u\left(y_{p_i}\right), \sum\nolimits_{p_i,p_j\in Y_{s_i}, p_i<p_j} \psi_p\left(y_{p_i}, y_{p_j}\right)\right). \tag{4}$$

Here, $\Omega^u$ represent a global weight and $\mathscr{C}$ represents a classifier, which identifies pixels as belonging to one of two segment types: complete or broken segments. The set of segments is represented by $S = S_c \cup S_b = \{s_1, \dots, s_n\}$, which is the union of two segment sets; complete segments, $S_c$, and broken segments, $S_b$, respectively. These sets are utilized to create segment cliques, which are groups of interconnected segments categorized into pair cliques, and completion and broken cliques. Sample cliques from each type are shown in Fig. 2a-c, where broken segments are represented in red and completion segments in yellow. The black pixels represent background pixels, gray pixels represent foreground pixels, and white pixels correspond to isolated points. These cliques represent different configurations of segments based on their relationship and interaction with each other. A pair clique, $C^P$, shown in Fig. 2a, is comprised of a pair of broken and completion segments. A completion clique, $C^C$, shown in Fig. 2b, describes a completion segment and its neighboring broken segments. Finally, a broken clique, $C^B$, shown in Fig. 2c, contains all completion segments connecting to either end of a broken segment.

At the classification stage, in addition to the broken and complete segments, triple junctions, isolated points, and the natural curving angles of segments are identified, which are shown in Fig. 2d for a sample section of the segmentation mask. In order to obtain completion segments, first all possible completion segments between all pairs of isolated points are generated, which is shown in Fig. 2e. Then, viable completion segments, which are defined as non-occluding completion segments below a predetermined length, are down selected from all possible segments, which are shown in Fig. 2f. Hence, the classifier maps the unary $\psi_u\left(y_{p_i}\right)$ and pairwise $\psi_p\left(y_{p_i}, y_{p_j}\right)$ energies of pixels constituting a segment, $s_i$, to the segment's unary energy $E_u(Y_{s_i})$, where $\psi_p\left(y_{p_i}, y_{p_j}\right)$ is given by:

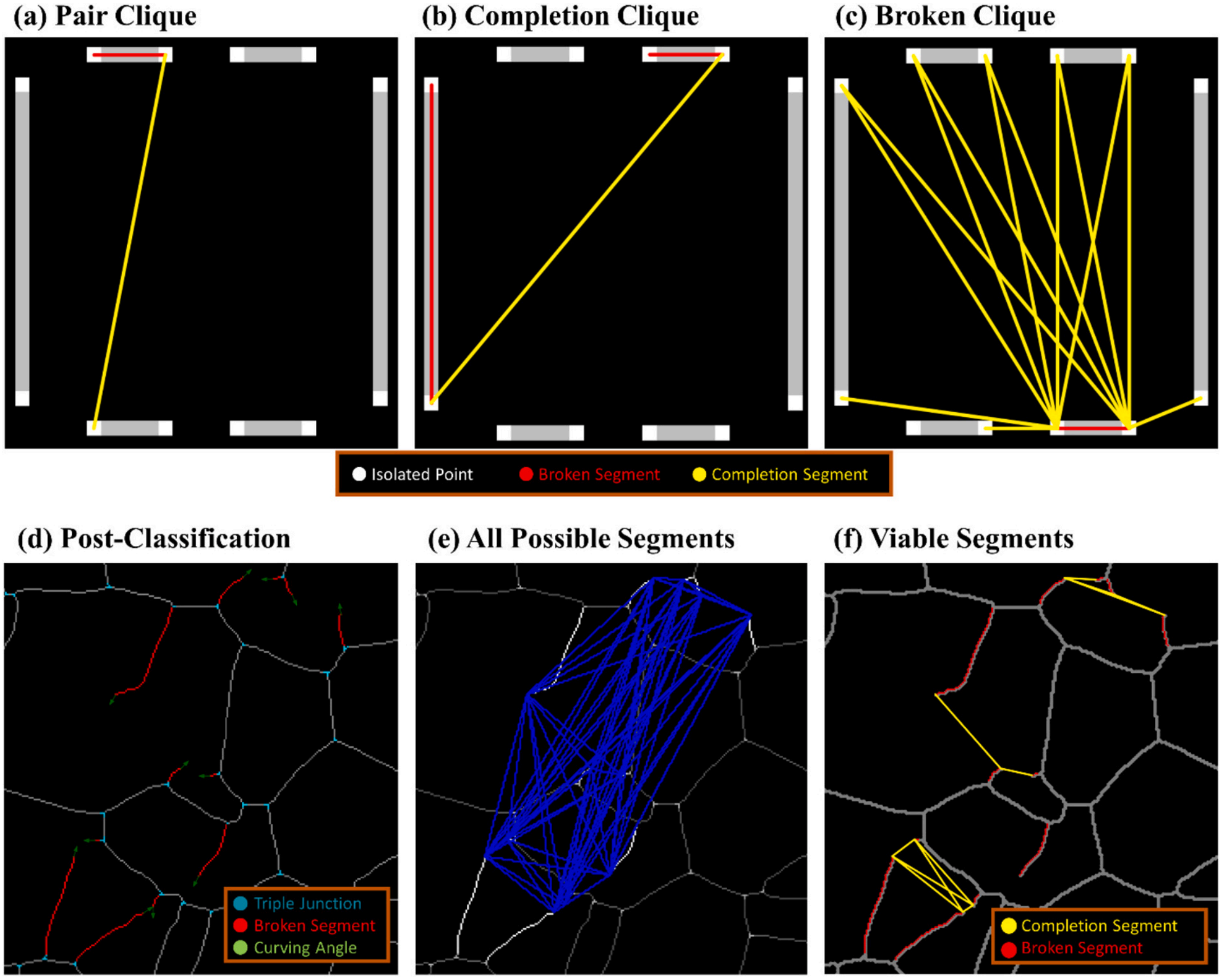

**Fig. 2. Visualization of cliques and classification processes. (a)** Pair Clique ($C^P$): Interaction between a broken (red) and completion (yellow) segment. **(b)** Completion Clique ($C^C$): A completion segment with its neighboring broken segments. **(c)** Broken Clique ($C^B$): Completion segments connecting to a central broken segment. **(d)** Post-classification detailing broken segments, triple junctions, and natural curving angles in a segmentation mask. **(e)** All possible completion segments generated between isolated points. **(f)** Selection of viable completion segments from the generated possibilities. (For interpretation of the references to color in this figure legend, the reader is referred to the web version of this article.)

$$\psi_p\left(y_{p_i}, y_{p_j}\right) = \mu\left(y_{p_i}, y_{p_j}\right)\sum_{m=1}^{K}\left[\omega_p{}^m k_p{}^m\left(f_i, f_j\right)\right], \tag{5}$$

where the pairwise-pixel features are given by the following equations:

$$k_p{}^{(1)}\left(f_i, f_j\right) = \omega_p{}^{(1)}\exp\left\{-\left[\frac{\left|p_i - p_j\right|^2}{2\theta_a^2} + \frac{\left|I_i - I_j\right|^2}{2\theta_b^2}\right]\right\}, \text{and}$$

$$k_p{}^{(2)}\left(f_i, f_j\right) = \omega_p{}^{(2)}\exp\left\{-\left[\frac{\left|p_i - p_j\right|^2}{2\theta_c^2}\right]\right\}. \tag{6}$$

Here, $\mu\left(y_{p_i}, y_{p_j}\right)$ is a label compatibility function that follows the Potts model for cost assignments, which states that the cost is non-zero only if the pixels are labeled differently. In other words, $\mu\left(y_{p_i}, y_{p_j}\right) = 1$ if $\left(y_{p_i} \neq y_{p_j}\right)$, and 0 otherwise. For a binary segmentation task, this model punishes similar pixels with different labels. Pairwise terms are calculated using two features involving Gaussian and bilateral filters, which are instrumental in satisfying the similarity and proximity principles. These principles suggest that pixels that are closely related in intensity and proximity tend to be grouped together. $\theta_a$, $\theta_b$, and $\theta_c$ are the parameters for the Gaussian and bilateral filters that can be adjusted, and $\omega_p{}^{(1)}$ and $\omega_p{}^{(2)}$ are the corresponding local weights.

The significance of closure principle in visual perception, while challenging to encapsulate through local potential functions, is approximated in the study by Ming et al. [79] via a series of localized connectedness constraints. Within a segment's terminal point, two kinds of connectedness constraints are defined, each defined with a series of linear inequalities:

- *The Completion Constraint* is characterized by the limitation that a completion segment can only be activated if its adjacent broken segment is also active, which requires the following inequality to be satisfied: $y_{s_i} \leq y_{s_j} \forall (s_i \in S_c) \cup \left(s_j \in S_b\right) \in C^P$.
- *The Extension Constraint* is defined by the condition that when a broken segment is active, at least one neighboring completion segment must also be active for possible extension. This is expressed as the inequality: $y_{s_j} \leq \sum_{s_i} y_{s_i} \forall (s_i \in S_c) \cup \left(s_j \in S_b\right) \in C^B$.

These constraints can be represented in the pairwise term $E_p\left(Y_{s_i}, Y_{s_j}\right)$ for segments with the following expression:

$$E_p\left(Y_{s_i}, Y_{s_j}\right) = M\left(\sum_{(s_i \in S_c)\cup\left(s_j \in S_b\right)\in C^P}\left[\left(1 - y_{s_j}\right)y_{s_i}\right] + \sum_{(s_i \in S_c)\cup\left(s_j \in S_b\right)\in C^B}\left[y_{s_j}\prod\left(1 - y_{s_i}\right)\right]\right) \tag{7}$$

where $M$ is a large constant, which is large enough to ensure that Eq. 1 results in a zero-probability configuration (i.e., $P(Y|X) = 1/Z(X)\,exp\{-M\} \approx 0$) in the case that these inequalities are not satisfied.

In line with the law of good figure, which generally favors simpler, more coherent shapes, the higher-order term $\left(E_h\left(Y_{s_i}, Y_{s_j}, Y_{s_k}\right)\right)$ is defined to obtain the best possible label configuration that is grounded in the domain expertise. Similar to the models in Ref. [22,79], different potentials are utilized to define a higher-order Gibbs energy. In this work, the higher-order term is represented as the sum of an interface potential and a complexity potential:

$$E_h\left(Y_{s_i}, Y_{s_j}, Y_{s_k}\right) = \Omega^I\sum_{\left(s_i, s_j, s_k\right)\in C^C}\left[\psi_I\left(y_{s_i}, y_{s_j}, y_{s_k}\right)\right] + \Omega^M\sum_{(s_i \in S_c)\cup\left(s_j, s_k \in S_b\right)}\left[\psi_M\left(y_{s_i}, y_{s_j}, y_k\right)\right], \tag{8}$$

where $\Omega^I$ and $\Omega^M$ are global weights. The interface potential is defined as:

$$\psi_I\left(y_{s_i}, y_{s_j}, y_{s_k}\right) = \sum_{m=1}^{K}\left[\omega_I^m k_I^m\left(f_i, f_j, f_k\right)\right]y_{s_i}y_{s_j}y_{s_k}, \tag{9}$$

where $\omega_I^m$ are the local weights for $k_I^m\left(f_i, f_j, f_k\right)$, which represent the interface features. There are seven interface features, the first four of which are obtained from Ref. [79], and an additional three are considered in this work. The initial four features are: (1) effective corner distance, optimized for capturing sharp turns in contours; (2) effective smooth distance, designed for capturing smoother transitions between segments; (3) angular completion, emphasizes geometric shapes through the sum of angles; and (4) angular difference, useful for identifying parallel or nearly parallel segments. The final three features pertain to the lengths of the two broken segments the one completion segment that forms the clique. Finally, the complexity potential is given by:

$$\psi_M\left(y_{s_i}, y_{s_j}, y_k\right) = \sum_{m=1}^{K}\left[\omega_M^m k_M^m\left(f_i, f_j, f_k\right)\right]y_{s_i}y_{s_j}y_{s_k}, \tag{10}$$

where $\omega_M^m$ are the local weights for $k_M^m\left(f_i, f_j, f_k\right)$, which includes two features representing the global complexity. These two features are: (i) $k_M^{(1)}$, which is the total effective length of the connected segments, and (ii) $k_M^{(2)}$, which represents the angle deviation from the ideal curving angle and the connection angle, effectively controlling the model's overall complexity.

The constraints introduced in Eq. 7 posit that $y_{s_j}$ and $y_{s_k}$ can both be active at the same time if and only if $y_{s_i}$ is also active, therefore simplifying the higher-order term to:

$$E_h\left(Y_{s_i}, Y_{s_j}, Y_{s_k}\right) = \Omega^I\sum_{\left(s_i, s_j, s_k\right)\in C^C}\left[\sum_{m=1}^{K}\left(\omega_I^m k_I^m\right)\right]y_{s_i} + \Omega^M\sum_{(s_i \in S_c)\cup\left(s_j, s_k \in S_b\right)}\left[\sum_{m=1}^{K}\left(\omega_M^m k_M^m\right)\right]y_{s_i}. \tag{11}$$

which in turn allows framing the combinatorial optimization problem as a MILP problem, for which the final energy function to be minimized takes the form:

$$min\left[\Omega^u\,\mathscr{C}\left(\sum_{p_i \in Y_{s_i}}\psi_u\left(y_{p_i}\right), \sum_{p_i, p_j \in Y_{s_i}, p_i < p_j}\mu\left(y_{p_i}, y_{p_j}\right)\sum_{m=1}^{K}\left[\omega^m k^m\left(f_i, f_j\right)\right]\right) + \Omega^I\sum_{\left(s_i, s_j, s_k\right)\in C^C}\left[\sum_{m=1}^{K}\left(\omega_I^m k_I^m\right)\right]y_{s_i} + \Omega^M\sum_{(s_i \in S_c)\cup\left(s_j, s_k \in S_b\right)}\left[\sum_{m=1}^{K}\left(\omega_M^m k_M^m\right)\right]y_{s_i}\right], \tag{12}$$

subject to $y_{s_i} \leq y_{s_j} \forall (s_i \in S_c) \cup \left(s_j \in S_b\right) \in C^P$ and $y_{s_j} \leq \sum_{s_i} y_{s_i} \forall (s_i \in S_c) \cup \left(s_j \in S_b\right) \in C^B$. Details pertaining to the implementation of pixel-based and segment-based CRFs are provided in Supplementary Notes 4 and 5, respectively.

### 2.4. *Tracing smooth paths*

Building on the probabilistic image labeling achieved through the

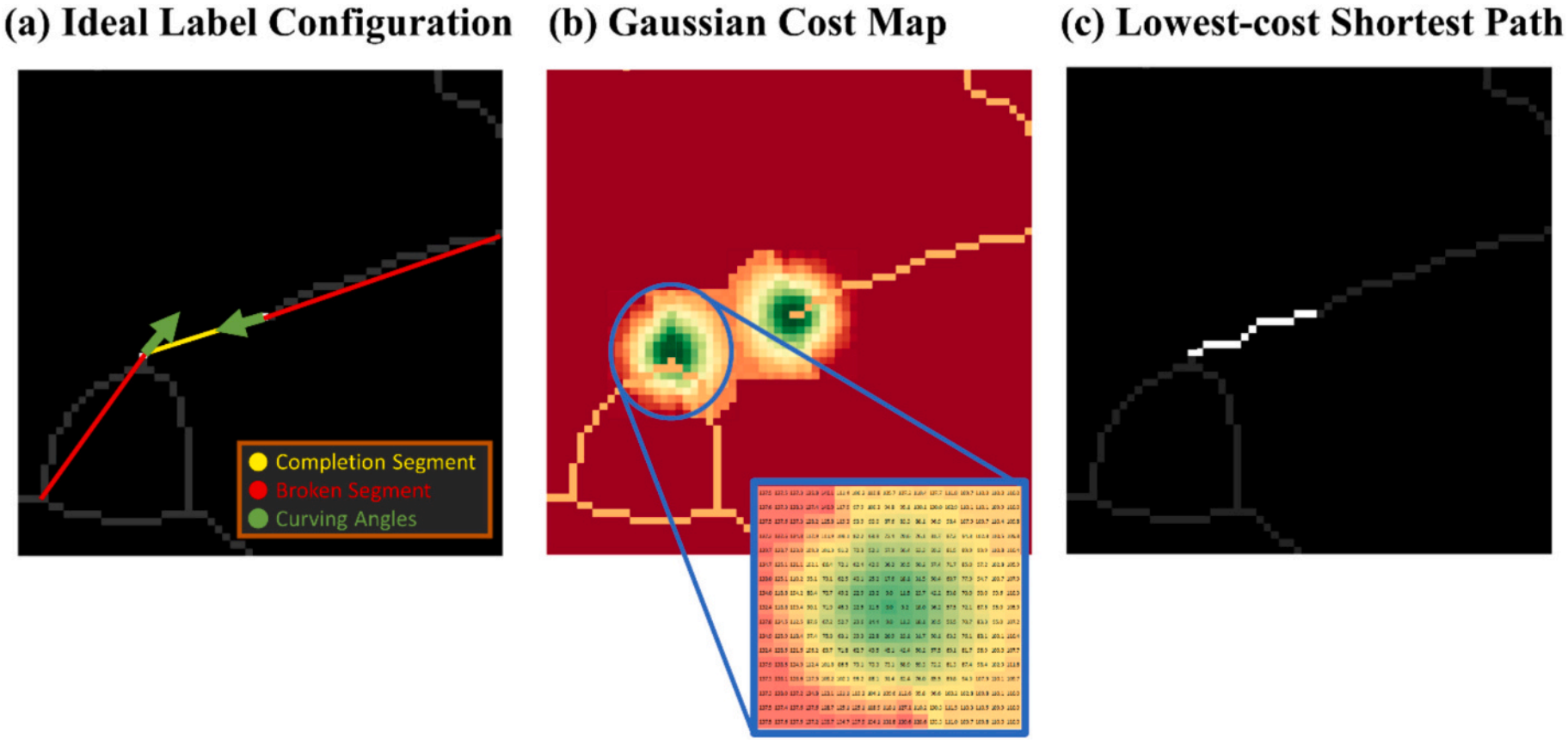

**Fig. 3.** A cost-based pathfinding algorithm. (a) The ideal label configuration with identified broken and completion segments, alongside curving angles. (b) The Gaussian cost map with varying color intensities and specific cost values. The inset illustrates the different costs assigned to pixels drawn from a 2D exact Gaussian distribution, based on the curving angle. (c) The lowest-cost shortest path is determined post-application of the A* pathfinding algorithm.

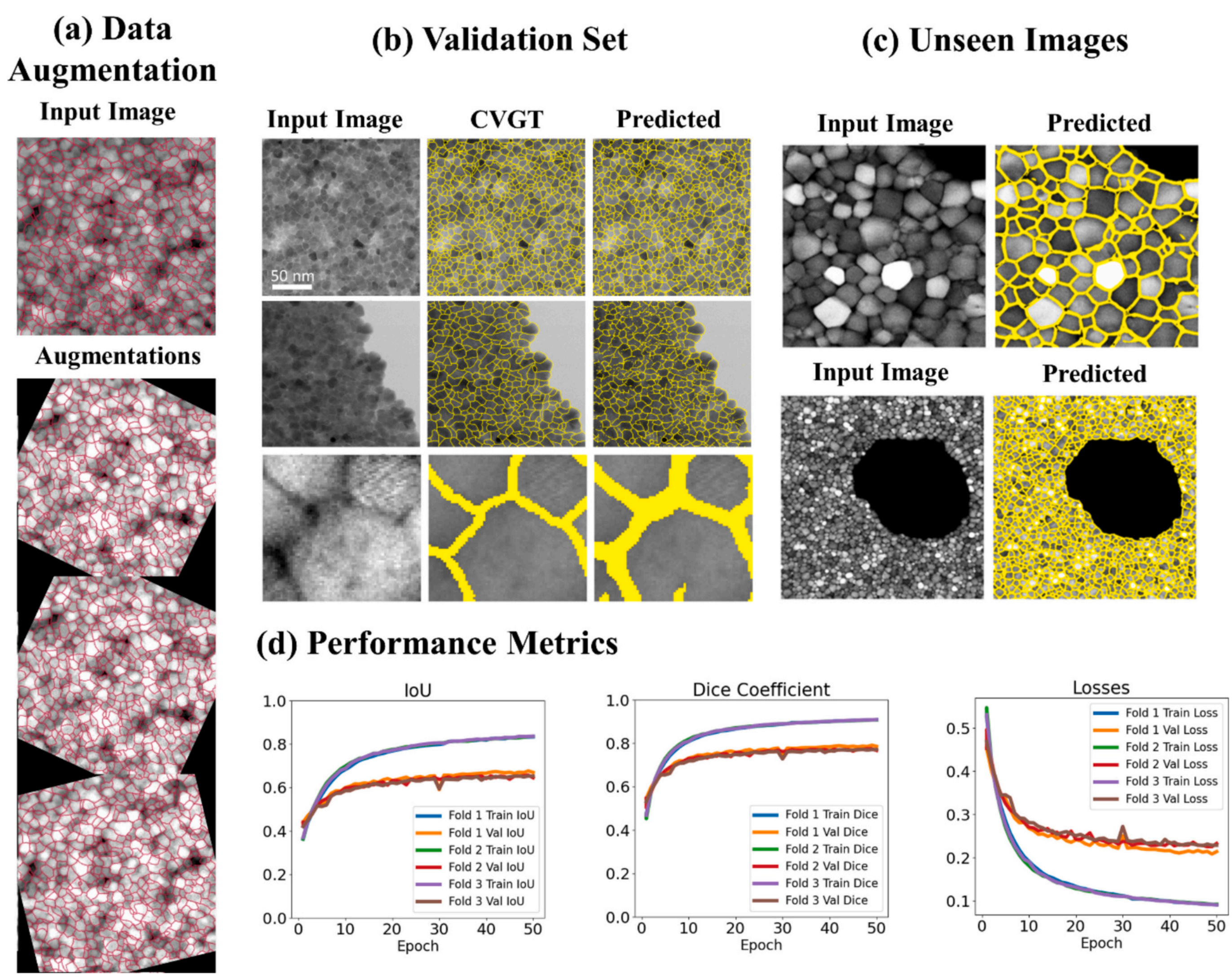

**Fig. 4. Computer vision model training and performance. (a)** One of the original input images and its corresponding data augmentations. (**b**) Images from the validation set, highlighting the computer vision ground truth (CVGT) and model-predicted fragmented segmentation masks. (**c**) The model's predictions on unseen images. (**d**) Plots of the intersection-over-union (IoU) and dice similarity coefficient (DSC) metrics along with the loss function (1-DSC) across training epochs for each fold.

Hierarchical CRF method, the next step in ensuring seamless connectivity involves tracing smooth, efficient paths. Existing methods like the stochastic completion fields method [59,76] can trace smooth paths between points, yet they lack guaranteed solutions and can be computationally intensive. Hence, in this work, where the identification of GBs for subsequent chemical analysis is time-sensitive, a cost-based pathfinding algorithm is utilized to find the lowest-cost shortest path between the endpoints of a completion segment, denoted as the source and the sink. Cost maps are generated using isolated points and curving angles of each segment in the ideal label configuration, as shown in Fig. 3a. Costs are derived from probability distributions based on the natural curving angles of adjacent broken segments (Fig. 3b), aiding in connectivity assumptions at interfaces. This method mirrors GBs' energy-minimization tendency, favoring shorter, less costly paths similar to how GBs naturally form by reducing their area (length in a projection) to lower their energy state. The selected pathfinding algorithm is the A* algorithm, a graph traversal and path search algorithm [32] that employs a heuristic function, here chosen to be the Chebyshev distance with 8-way directionality, with the rationale for this selection being elaborated in Supplementary Note 1. This algorithm starts from the source, moving towards the sink while tracking path and associated cost. Paths are reconstructed by backtracking the lowest-cost shortest paths identified by A*, as in Fig. 3c. These paths, shown in Fig. 3g, connect broken segments in fragmented masks, resulting in refined segmentation depicted in Fig. 3h, reflecting GBs' natural energy-minimizing behavior. Further details on the pathfinding algorithm can be found in Supplementary Note 1.

## 3. Results

To demonstrate this methodology, fragmented segmentation masks are generated using the trained computer vision algorithm. The annotated images serve as ground truth for model training, labeled as CVGT to differentiate from fragmented segmentation ground truth (FSGT). CVGT assesses the computer vision model's performance, while FSGT evaluates segmentation refinement. To enhance the robustness and generalizability of the model, data augmentation methods were carefully selected. Brightness adjustment is used to simulate variations in illumination, Gaussian blur to mimic slight focus variations common in microscopy, and rotation to ensure the model's insensitivity to orientation changes. These augmentations aim to produce a more versatile and adaptable computer vision algorithm. A sample of these augmentations is displayed in Fig. 4a. The trained model, described in the Methods section, is then utilized to generate segmentation masks from the input images. The model effectively delineates boundaries in both validation and unseen images, as shown in Figs. 4(b) and (c), and in comparisons between CVGTs and predicted masks. It successfully generates segmentation masks for images not included in the training, performing better on images with fewer atomic features and pronounced contrast.

In this study, a 3-fold cross-validation strategy was employed during network training over 50 epochs. The aggregated metrics across folds indicate an average training IoU of approximately 0.83 and a corresponding validation IoU near 0.67, while the training DSC approached 0.91 and the validation DSC stabilized at around 0.79, as shown in Fig. 4d. These similar performance values across folds reflect that the model is reliably capturing the key features of the grain boundaries and generalizing well to unseen data. Specifically, the consistency of these metrics across different splits suggests that the model is not overly sensitive to the specific partitioning of the data, which is a strong indication of robust learning rather than overfitting.

In addition, an alternative training approach was explored that consolidated all training data and extended the training duration. Although this method yielded higher training scores—a result of the more intensive training—the validation scores remained comparable to those of the cross-validation model. While both approaches deliver similar validation performance, the cross-validation model confirms the robustness of the segmentation pipeline, and the alternative approach is particularly valuable for demonstration purposes because it produces sharper, more detailed segmentation outputs. This enhanced detail facilitates a clearer presentation of the efficacy of the subsequent post-processing stage in rectifying fragmented masks.

The performance of the model aligns with benchmarks in thin segmentation mask tasks, comparable to a DSC of 0.7 [34] reported by Jahangard et al. for retinal blood vessel segmentation. Given the focus of this work on refining the fragmented segmentation mask, further optimizations like hyperparameter tuning are not pursued. The model's predictions on full-resolution images, critical for analyzing detailed material structures potentially lost or distorted in lower resolutions, demonstrate its efficiency in processing high-resolution images.

On the fragmented segmentation masks, pixel-based CRFs are applied, followed by the creation of the FSGT. FSGT is formed by manually labeling each identified completion and broken segment, following specific principles including consistent judgement across all images, marking the connection segments only if the natural curvatures of the broken segments appear to intersect and ensure non-occluding connections. These FSGTs aid in learning optimal weights and evaluating the proposed method's performance. In a sample image, Fig. 5a displays all segments identified by the algorithm, Fig. 5b the manually labeled FSGT, and Fig. 5c the best solution from the differential evolution algorithm, which corresponds to the minimum difference in labeling configurations between the FSGT and MILP solution. This accuracy is calculated as the percentage of correctly classified segments (i.e., true positives and true negatives) from the total number of viable segments. Insets in Fig. 5 magnify portions where the CRF accurately predicts most segments, though as noted, some discontinuities may persist where the probabilistic evidence for a connection is insufficient. Additional magnified views are shown in Fig. S2. Fig. 5c illustrates newly identified grains from connected segments, enhancing segmentation accuracy, as the subsequent watershed algorithm identifies grains by continuous lines forming the boundary of closed geometric shapes.

### 3.1. *Watershed segmentation*

A marker-based watershed segmentation technique, based on geographical watershed principles [16,62], is employed for image segmentation. This method converts the image into a topographical map with pixel values representing altitudes, applying the watershed transform to create catchment basins associated with distinct markers. The basin boundaries, as shown in Fig. 6a, segment the image into distinct regions or grains. However, broken lines in the mask, such as those visible in Fig. 6a, can result in suboptimal segmentation. The grains in these plots appear as randomly colored shapes, except for the region depicted in the insets, where grain colors are matched manually for better visualization. More details about this method can be found in Supplementary Note 2. This segmentation is applied to three image types: (1) the original segmentation mask before post-processing (Fig. 6a), (2) the FSGT (Fig. 6b), and (3) the CRF solution (Fig. 6c). The figure shows that broken segments without viable completion segments are removed, and sections are smoothly connected, thereby enhancing segmentation. Insets in Fig. 6 demonstrate how the connections shown in Fig. 5c results in better grain segmentation, impacting the statistical representation of GB networks through improvements in grain shape and area. Additional magnified views are shown in Fig. S3. These segmented images are then used to evaluate the performance of the post-processing procedure.

### 3.2. *Performance evaluation*

In the segmentation performance evaluation, the post-processing procedure consistently improved various metrics, as shown in Table 1. Pixel accuracy, measuring correctly classified pixels against the total number, increased by 0.18 % in the validation set from an already high baseline of 0.997 to 0.999, indicating near-perfect classification accuracy. To better assess segmentation masks, metrics such as IoU and DSC are often employed [19]. IoU and DSC metrics exhibited an increase of 9.15 % and 5.19 % in the validation set due to post-processing, respectively. These improvements bring the final IoU and DSC values to 92.5 % and 96.1 % representing significant enhancements in segmentation accuracy, particularly for tasks involving complex grain structures. Calculation details for these metrics are provided in Supplementary Note 3. Additionally, grain-specific metrics, like the number of grains, $\boldsymbol{N_G}$, and average grain size, $\boldsymbol{V_G}$, also demonstrated notable improvements, detailed in Table 2. These enhancements were calculated as percentage improvements over the FSGT baseline, based on their relative differences.

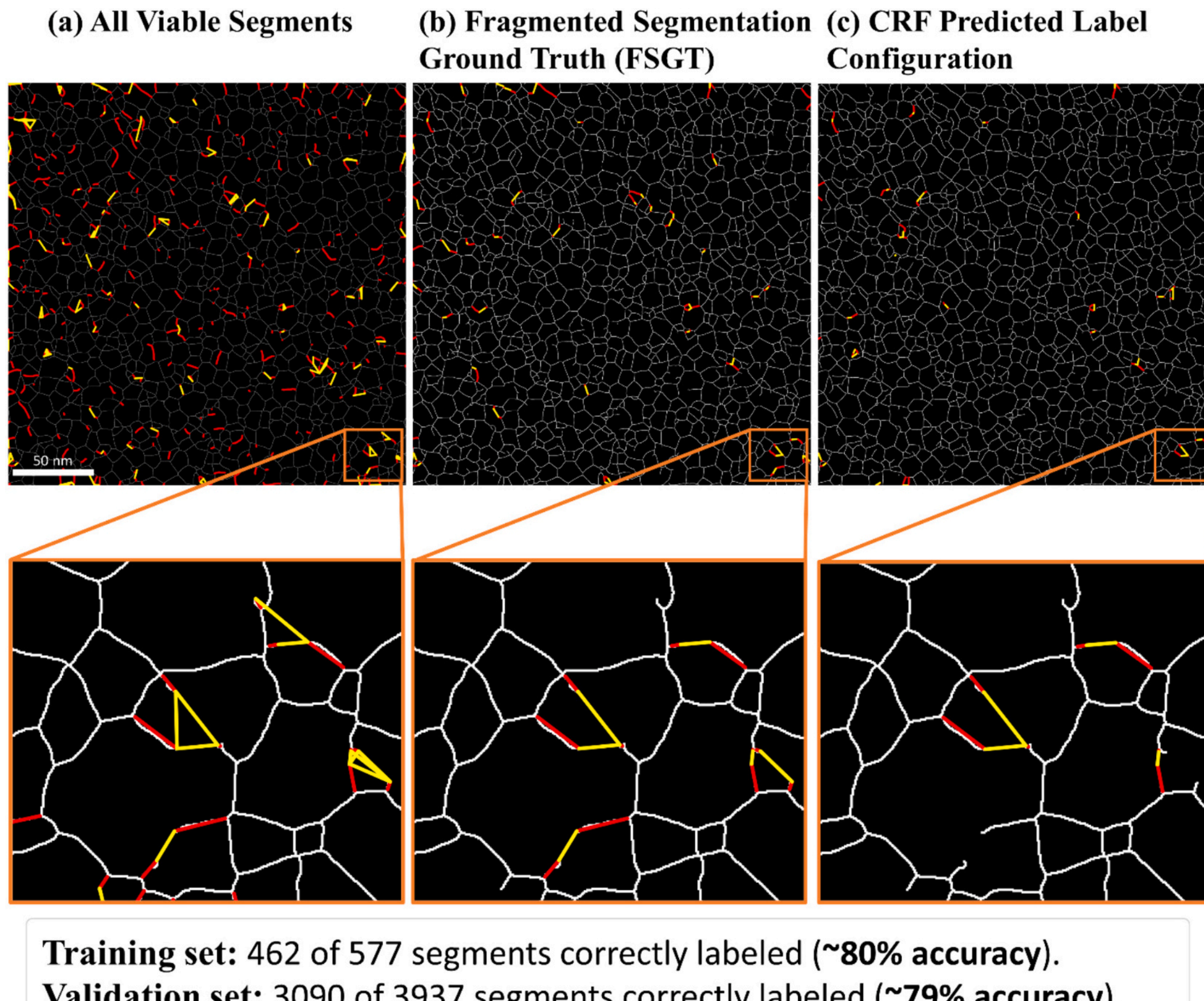

**Fig. 5. Detailed examination of the segment-based CRF labeling. (a)** All viable segments identified by the algorithm with broken (red) and completion (yellow) segments. (**b)** The manually labeled fragmented segmentation ground truth (FSGT). (**c)** The CRF predicted label configuration, illustrating the optimal segment connections post-differential evolution algorithm. Insets in each subplot offer a magnified view of selected areas, elucidating the finer details of segmentation. The image scale is consistent with that shown in Fig. 4b. (For interpretation of the references to color in this figure legend, the reader is referred to the web version of this article.)

An analysis of confusion matrices is presented in Fig. 7a for the training set and Fig. 7b for the validation set. In the training dataset, the CRF correctly predicts 462 out of 577 segments, achieving approximately 80 % accuracy. For the validation set, it accurately predicts 3090 out of 3937 segments, a 79 % success rate. Regarding time efficiency, manual labeling to create the FSGT is a necessary but time-intensive prerequisite, taking roughly 41 min per 1000 segments. In contrast, the automated CRF method labeled 1000 segments in about 15 milliseconds on an Intel Core i7-12700K 12-Core Processor. This five-order-of-magnitude speed-up is not intended to diminish the crucial role of manual annotation for training and validation, but rather to highlight the framework's potential for high-throughput analysis where such manual intervention is infeasible. Additionally, weight learning, needed once per application, completes in around 47 min. For fully automated STEM imaging and spectroscopic data acquisition in GB network chemical analysis, the CRF's rapid processing rate would enable time-efficient implementation needed to mitigate the effects of microscope instability such as specimen drift and lens defocus.

A novel metric for grain alignment is developed to overcome the limitations of traditional metrics in this specific segmentation context. This metric computes the average absolute difference between the areas of closely aligned regions in two labeled images, identified based on the proximity of their centroids. A smaller value for grain alignment indicates a higher degree of similarity in terms of grain size between the two images, making it an effective measure for comparing segmented regions, especially in the context of thin segmentation masks. Significant improvements in mean grain alignment, up to 78 % in training and 51 % in validation sets, are observed post-segmentation. These enhancements are critical as they significantly enhance the accuracy of statistical representation of grain size and location, leading to more precise analysis of GB networks, especially in densely packed or intersecting boundary regions.

## 4. Discussion

### 4.1. Beyond conventional metrics

Pixel accuracy focuses on the count of foreground and background pixels and neglects their spatial distribution, which limits its utility in single-pixel wide masks. In cases where discarded broken segment pixels

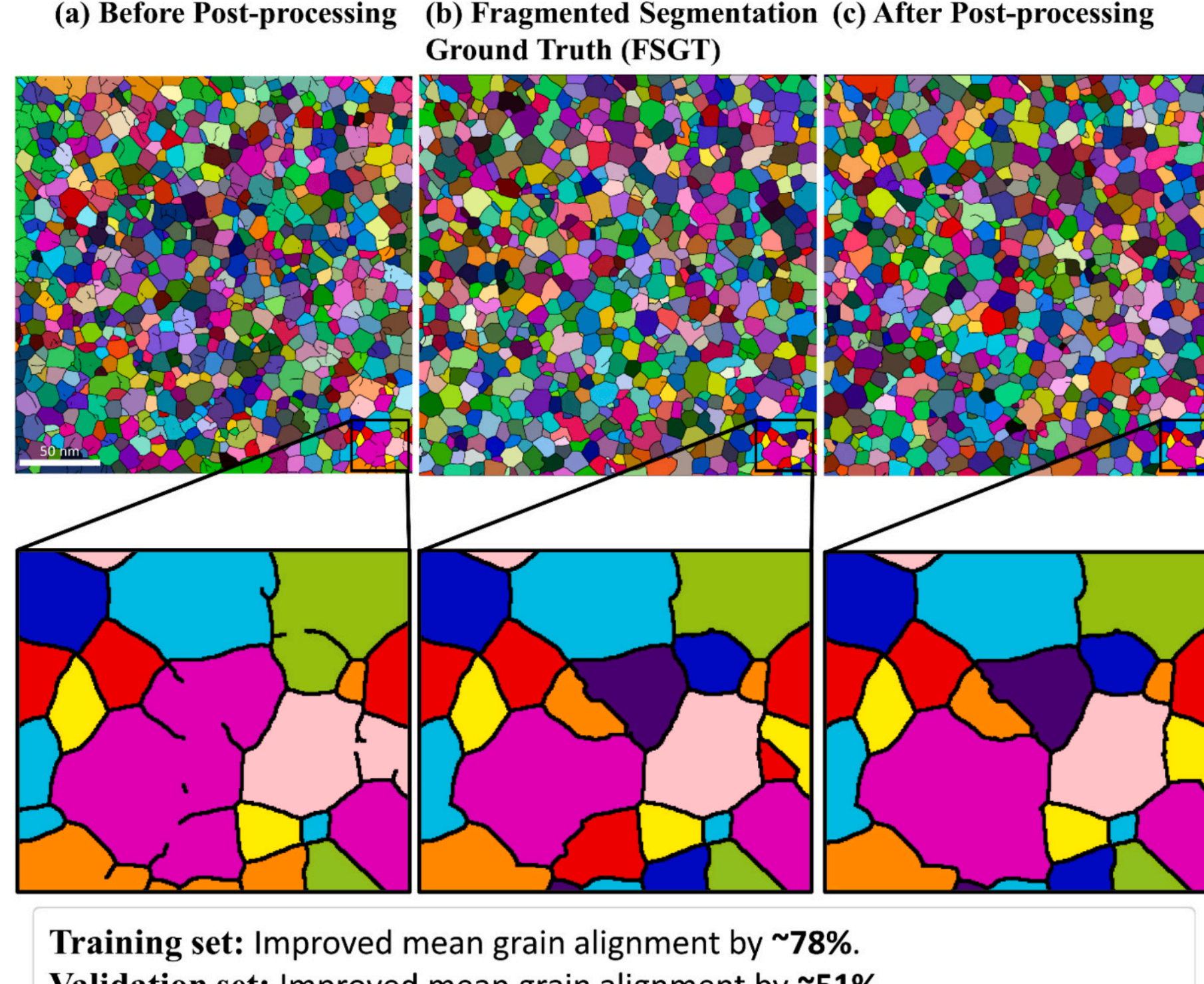

**Fig. 6. Image segmentation using marker-based watershed technique. (a)** Original segmentation mask before post-processing with visible broken segments. **(b)** Fragmented segmentation ground truth (FSGT) image that highlights the connections between the disconnected segments. **(c)** CRF solution corresponding to the optimal label configuration after post-processing, showing removal of dangling broken segments and smoother connections between sections. The grains are randomly colored, except for the insets. Insets in each subplot provide a magnified view, emphasizing the differences in grain identification and segmentation performance. The image scale is consistent with that shown in Fig. 4b.

**Table 1**
Comparison of segmentation performance evaluation metrics for both training and validation sets, highlighting the improvements achieved through post-processing.

| Metric | Training set | | | Validation set | | |
|---|---|---|---|---|---|---|
| | FSGT /before* | FSGT /after* | Percent change | FSGT /before* | FSGT /after* | Percent change |
| Accuracy | 0.998 | 0.999 | **0.18 %** | 0.997 | 0.999 | **0.18 %** |
| IoU | 0.911 | 0.979 | **6.78 %** | 0.833 | 0.925 | **9.15 %** |
| DSC | 0.954 | 0.989 | **3.58 %** | 0.909 | 0.961 | **5.19 %** |
| Precision | 0.929 | 0.996 | **6.66 %** | 0.853 | 0.967 | **11.36 %** |
| Recall | 0.979 | 0.983 | **0.39 %** | 0.973 | 0.955 | **−1.77 %** |
| Grain alignment | 777.8 | 174.2 | **77.6 %** | 4062.7 | 2007.7 | **50.6 %** |

The bold values indicate the percent changes in the metrics.
* Metrics calculated by comparing the FSGT and the segmentation mask before or after post processing.

**Table 2**
Metrics specific to grain attributes for both training and validation sets, illustrating the number of grains, $N_G$ and average size of identified grains, $V_G$ and their respective percent changes after post-processing.

| Metric | Training set | | | | Validation set | | | |
|---|---|---|---|---|---|---|---|---|
| | FSGT* | Before* | After* | Percent change | FSGT* | Before* | After* | Percent change |
| $N_G$ | 904 | 797 | 893 | **10.62 %** | 1518 | 1206 | 1520 | **20.69 %** |
| $V_G$ | 4309.0 | 4812.8 | 4429.2 | **8.90 %** | 10,864.0 | 13,293.4 | 11,180.7 | **19.45 %** |

The bold values indicate the percent changes in the metrics.
* Metrics calculated for different segmentation masks, including the FSGT and the segmentation mask before and after post processing.

equal the number of added completion segment pixels, pixel accuracy remains unchanged, potentially misrepresenting changes in spatial distribution. While high pixel accuracy may suggest superior performance, it can obscure spatial distribution nuances. IoU and DSC, although more nuanced, are sensitive to minor spatial variations, which is a notable limitation in thin masks. Nevertheless, improvements in these metrics post-processing affirm their utility in assessing the alignment of segmentation masks with ground truth.

Table 2 shows post-processing enhancements in actual segmented grains. Both the number of identified grains and average grain size align more closely with the ground truth post-processing, signifying not only improved alignment but also enhanced grain segmentation attributes.

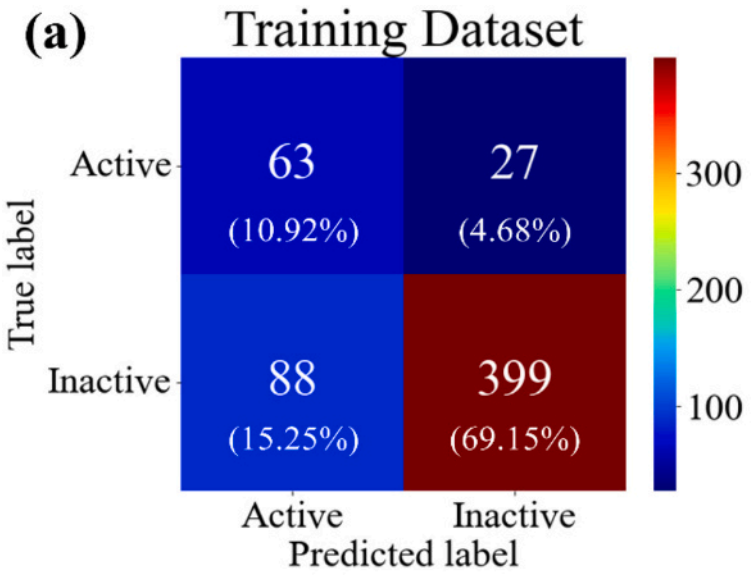

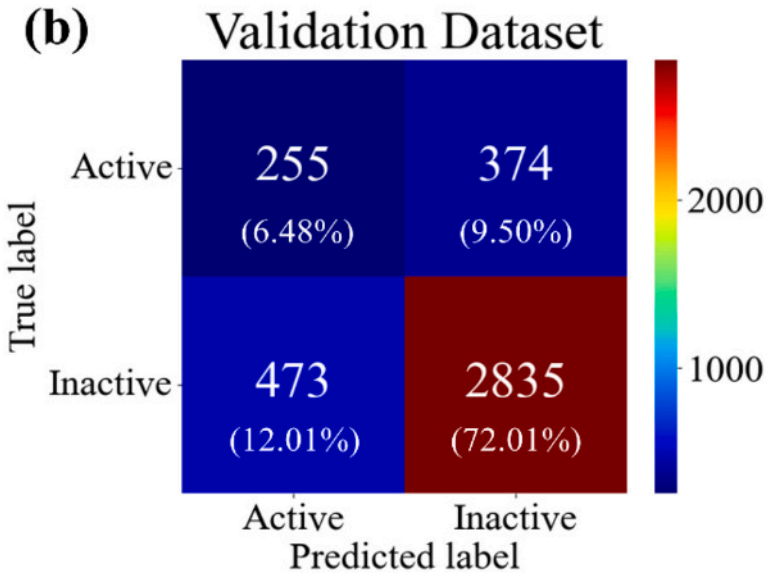

**Fig. 7. Detailed evaluation metrics for segmentation performance.** Confusion matrices illustrating the alignment of predicted segment labels with true labels for (**a**) the training set, and (**b**) the validation set. 'active' and 'inactive' denote segments identified as kept or discarded, respectively, highlighting the model's classification efficacy.

These advancements indicate a more accurate representation of grain size distribution and density, crucial for assessing mechanical and physical material properties. GBs considerably affect aspects like crack propagation, corrosion resistance, and mechanical strength, highlighting the importance of these improvements.

Table 1 and the confusion matrices in Fig. 7a and b reveal a decrease in recall for the validation set, suggesting the model's reduced sensitivity in identifying active segments post-processing. This could stem from unseen data or minor overfitting during training. These metrics primarily assess the model's proficiency in correctly labeling pixels or segments as active or inactive. However, in grain segmentation tasks that emphasize not only accurate pixel classification but also the morphological characteristics of the grains, such metrics may not fully capture the performance.

Addressing limitations in conventional metrics, a new metric, grain alignment, is proposed for evaluating segmentation performance. This metric involves identifying centroids of corresponding regions in two labeled images. For each grain in the first image, the closest centroid grain in the second image is determined, ensuring spatial proximity-based comparison. The absolute differences in the areas of these closely aligned regions are calculated and averaged to gauge size similarity, offering a robust measure for segmented region accuracy, particularly in thin masks. While this approach provides a robust metric, it's important to note the mean grain alignment metric difference between the training and validation dataset can be attributed to a few considerations. This discrepancy arises because accurately identifying grains requires all constituting lines to be connected. In fragmented images, unconnected lines can lead to incorrect grain identification, merging multiple grains into a single large one, and thus affecting the grain count and spatial distribution. The post-processing method rectifies this by connecting fragmented boundaries; by successfully closing a previously open boundary with a completion segment, a single, incorrectly merged grain is correctly partitioned into two or more distinct grains. This partitioning effect is the primary reason for the observed increase in the number of grains ($\boldsymbol{N_G}$) post-processing (Table 2), bringing the count closer to the ground truth. Additionally, the post-processing's single-pass might leave some segments incomplete. This difference is more indicative of the validation set's complexities rather than a general limitation of the method. By identifying closely situated grains as part of the same boundary network, this method precisely represents microstructures, essential for accurate material behavior prediction, demonstrating its utility despite the noted dataset-specific challenges.

In addition to the demonstrated accuracy, the CRF's remarkable time efficiency is particularly valuable in materials science applications requiring real-time analysis. This rapid segmentation capability, in stark contrast to the time-consuming manual labeling process, allows for immediate GB identification post-image acquisition. Crucial in countering issues like drift in TEM imaging, real-time processing ensures analyses are based on accurately segmented, current images. This not only boosts analysis reliability but also enables dynamic, in-situ experimentation and observation in materials science, where monitoring changes over time and under varying conditions is essential.

### 4.2. Comparing standard and specialized ground truths

The CVGT and FSGT evaluate distinct aspects of segmentation, targeting specific criteria. The CVGT focuses on essential features for initial computer vision training, whereas the FSGT addresses specialized grain segmentation requirements, including unique post-processing needs such as continuous grain lines. Consequently, comparing post-processed results directly with the CVGT is inherently flawed. The CVGT, serving as a baseline for assessing computer vision algorithm performance, varies across images and is contingent on the computer vision output's quality. Therefore, it becomes an unstable and unreliable baseline for evaluating the post-processing procedure's performance, which inherently refines and alters the computer vision algorithm's initial segmentation.

Post-processing impacts metrics like IoU and DSC, influenced by the initial model's alignment with the CVGT. Generally, good alignment enhances these metrics post-processing, whereas poor alignment diminishes them. For instance, in Table 1, using the CVGT as ground truth, IoU and DSC in the training set decrease by 0.12 % and 0.21 %, respectively, despite a 78 % improvement in grain alignment. This indicates that post-processing accentuates the initial segmentation's adherence to the CVGT. Therefore, to isolate the specific influence of the post-processing procedure, a comparison with the FSGT is essential. Additionally, the subjectivity in annotating GBs complicates evaluations. Consistency in CVGT and FSGT annotations is difficult due to specific grain characteristics, introducing variability in ground truths and affecting metrics. This underlines the limitations of relying solely on CVGT for evaluation and the importance of including the FSGT for comprehensive assessment.

### 4.3. Towards advanced segmentation and cross-domain applications

Factors like hyperparameter optimization, model choice, and more training data can enhance baseline segmentation accuracy. Post-processing adjustments involve hyperparameters like maximum segment distance for selecting viable segments, field size, and path-finding costs. Iteratively optimizing these hyperparameters is recommended for improved robustness and accuracy. This work also paves the way for joint CRF and CNN training. End-to-end trainable models, such as those that model CRFs and their feature functions as recurrent neural networks, can be appended to CNNs for training [83], which are expected to yield even higher segmentation accuracies [8].

In this study, T-junctions—connections between isolated points and segments—are not included, despite constituting about half of the training set's segment candidates. Their inclusion is expected to improve segmentation performance [79], offering a fuller network representation. Beyond the seven interface features examined in this study, future iterations could integrate additional grain-related features, such as the Mullins grain growth model [48] to inform the algorithm about potential GB movements for enhanced predictive accuracy. Combining electron diffraction with optical reflectance measurements to estimate GB character [64], and incorporating crystallographic orientation data, can provide detailed grain physical properties, aiding in segmentation refinement. Electron diffraction and optical reflectance could enhance GB identification, while crystallographic orientation might help differentiate grain types or phases. Further segmentation enhancement might come from local neighborhood analysis using image features like brightness, color, texture [47], and the intensity gradient along completion segments [20]. When the original image is accessible, contrast differences between grains and texture analyses could be employed between the original and the predicted masks [28]. This can link GB microstructure with material properties, aiding in predicting mechanical, thermal, and electrical behaviors [15].

Implementing this post-processing procedure facilitates high-throughput experimental pipelines with reduced human supervision requirements, while maintaining comparable accuracies. S(TEM) images can be subjected to computer vision algorithms and subsequently analyzed using complementary methods such as electron diffraction and spectroscopies in a continuous pipeline. Fractional probabilities obtained from computer vision outputs can serve as unary terms in pixel-based CRFs, enhancing predictions and enabling confidence level assignments to each segment for targeted experiments [39].

Statistical distribution information, such as average grain size, could be obtained and leveraged to improve segmentation performance. Outlying grains, indicative of abnormal growth, could be flagged for operator review. The post-processing procedure makes such data available to the operator, alongside candidate segments and their associated costs, thus facilitating a human-in-the-loop approach while expediting the process without compromising fidelity. The segment-based CRF approach exhibits strength in detecting closed contours in noisy environments [79], a capability that can be harnessed to identify interfaces even in the presence of other image features like nano-precipitates or atomic details from S(TEM) without an additional classifier [42].

Furthermore, the general applicability of this framework extends to less-ideal imaging conditions often encountered in experimental settings, such as in-situ microscopy. While the current work utilized high-quality STEM images, the post-processing method is fundamentally designed to correct fragmented masks, which are characteristic of noisy or low-contrast images. Similarly, for thicker samples or alloys exhibiting complex contrasts from features like segregation-induced complexions, the initial CNN segmentation may be challenged. However, the CRF-based refinement, which operates on the geometric properties of the predicted mask, remains robust. It enforces structural priors like continuity and closure, helping to reconstruct the grain boundary network even when local pixel evidence is ambiguous. Future work could enhance this by incorporating image features specifically sensitive to such complex contrast variations into the CRF's potential functions.

While the presented work focuses on grain boundaries, the core methodology is adaptable to other domains. The strength of the hierarchical CRF lies in its general framework of applying perceptual grouping rules to refine fragmented line networks. The utility of thin segmentation masks extends beyond the domain of materials science, finding applications in biomedical imaging for narrow or elongated structures. Areas of application include retinal blood vessel segmentation [34], neuronal structure analysis [33], coronary artery delineation in cardiac MRI [17], tendon segmentation in musculoskeletal imaging [35], lymphatic system studies [41], and angiography for small blood vessels [82]. However, the specific feature functions and learned weights are necessarily domain-specific. Applying this method to other problems, would require retraining on domain-specific ground truths and could benefit from designing features tailored to the unique characteristics of those images. The methodology introduced in this work offers a new perspective and serves as a resource for other researchers, setting the stage for further advancements in segmentation mask optimization for interconnected line networks.

## CRediT authorship contribution statement

**Doruk Aksoy:** Writing – review & editing, Writing – original draft, Software, Methodology, Investigation, Formal analysis, Data curation, Conceptualization. **Huolin L. Xin:** Writing – review & editing, Writing – original draft, Funding acquisition. **Timothy J. Rupert:** Writing – review & editing, Writing – original draft, Supervision, Funding acquisition. **William J. Bowman:** Writing – review & editing, Writing – original draft, Visualization, Supervision, Project administration, Methodology, Investigation, Funding acquisition, Data curation, Conceptualization.

## Code availability

The underlying code for this study is available in the HierarchicalCRF repository and can be accessed via this link: https://github.com/dorukaksoy/HierarchicalCRF.

## Declaration of competing interest

All authors declare no financial or non-financial competing interests.

## Acknowledgements

This research was primarily supported by the National Science Foundation Materials Research Science and Engineering Center program through the UC Irvine Center for Complex and Active Materials (DMR-2011967).

## Appendix A. Supplementary data

Supplementary data to this article can be found online at https://doi.org/10.1016/j.matchar.2025.115694.

## Data availability

All data generated or analyzed during this study are included in this published article and its supplementary information files. A preprint version of this manuscript is available on arXiv [7].